\documentclass[pdflatex,sn-mathphys-num]{sn-jnl}
\usepackage[latin9]{inputenc}
\usepackage{amsmath}
\usepackage{amssymb}
\usepackage{graphicx}

\makeatletter

\@ifundefined{textcolor}{}
{%
 \definecolor{BLACK}{gray}{0}
 \definecolor{WHITE}{gray}{1}
 \definecolor{RED}{rgb}{1,0,0}
 \definecolor{GREEN}{rgb}{0,1,0}
 \definecolor{BLUE}{rgb}{0,0,1}
 \definecolor{CYAN}{cmyk}{1,0,0,0}
 \definecolor{MAGENTA}{cmyk}{0,1,0,0}
 \definecolor{YELLOW}{cmyk}{0,0,1,0}
}

\usepackage{amsfonts}
\usepackage{graphicx}
\usepackage{graphics}
\usepackage[usenames]{color}

\makeatother

\begin{document}
\title{Hyperspherical Analysis of Dimer-Dimer Scattering in One-Dimensional Systems}
\author*[1]{\fnm{Jia} \sur{Wang}}\email{jiawanghome@gmail.com}
\author[1]{\fnm{Hui} \sur{Hu}}
\author[1]{\fnm{Xia-Ji} \sur{Liu}}
\affil[1]{\orgdiv{Centre for Quantum Technology Theory}, \orgname{Swinburne University of Technology},
\orgaddress{\city{Melbourne}, \postcode{3122}, \country{Australia}}}

\abstract{We present a comprehensive analysis of four-body scattering in one-dimensional (1D) quantum systems using the adiabatic hyperspherical representation (AHR). Focusing on dimer-dimer collisions between two species of fermions interacting via the sinh-cosh potential, we implement the slow variable discretization (SVD) method to overcome numerical challenges posed by sharp avoided crossings in the potential curves. Our numerical approach is benchmarked against exact analytical results available in integrable regimes, demonstrating excellent agreement. We further explore non-integrable regimes where no analytical solutions exist, revealing novel features such as resonant enhancement of the scattering length associated with tetramer formation. These results highlight the power and flexibility of the AHR+SVD framework for accurate few-body scattering calculations in low-dimensional quantum systems, and establish a foundation for future investigations of universal few-body physics in ultracold gases.}

\maketitle

\section{Introduction}

Since the groundbreaking development of laser cooling and trapping
of atoms \citep{Metcalf1999book}, the field of ultracold quantum
gases has rapidly expanded and made a profound impact on the study
of fundamental quantum phenomena \citep{Stringari1999RMP,Leggett2001RMP,Stringari2008RMP}.
At temperatures as low as a billionth of a degree Kelvin (nano-Kelvin,
nK) and densities ranging from $10^{12}$ to $10^{15}$ ${\rm cm}^{-3}$,
these systems reach the quantum degenerate regime, where the atomic
de Broglie wavelength exceeds the typical interparticle spacing \citep{Cornell1995Science,Jin1999Science}.
A variety of trapping techniques enable the realization of low-dimensional
systems, while Feshbach resonances allow for precise and tunable control
over interparticle interactions \citep{BlochRMP2008,ChinReview2010}.
This extraordinary level of control, combined with recent advances
in theoretical and computational techniques, makes ultracold gases
an ideal platform for exploring few-body quantum physics and performing
quantitative comparisons between theory and experiment.

One of the most striking few-body phenomena predicted in this context
is the Efimov effect. In the early 1970s, Vitaly Efimov predicted
that three identical resonant bosons can form a series of three-body
bound states---now known as Efimov trimers---characterized by a
discrete scaling symmetry \citep{Efimov1970YZ}. Remarkably, the scaling
factor is independent of short-range interaction details, making the
phenomenon universal. Once considered purely theoretical, Efimov states
have since been observed in numerous ultracold gas experiments \citep{Grimm2006Nature,Jochim2008PRL,Modugno2009NP,Hulet2009Science,Khaykovich2009PRL,OHara2009PRL1,OHara2009PRL2,Ueda2010PRL,Minardi2009PRL,Khaykovich2010PRL,Jochim2010PRL}.
Moreover, new forms of universality have emerged in these systems,
with Efimov trimers displaying properties that reflect the long-range
van der Waals tails of the underlying two-body potentials \citep{HutsonPRL2011,Jin2012PRL,Jia2012PRL,Yujun2012PRL,DIncao2013FBS}.
Other exotic few-body states, including super-Efimov \citep{Nishida2013PRL,Gridnev2014JPA,Jia2015PRA}
and semi-super Efimov states \citep{Nishida2017PRL}, have been theoretically
proposed and are being actively explored in ultracold gases, particularly
in systems with tunable dimensionality.

In this work, we investigate the four-body scattering problem in one-dimensional
(1D) systems. A system of four fermions---two in one component and
two in another---is the simplest nontrivial few-body configuration
that captures the essential physics of ultracold fermionic gases.
We numerically obtain essentially exact solutions to this problem
using the adiabatic hyperspherical representation (AHR). AHR is a
well-established method in atomic, molecular, and optical physics,
as well as in nuclear and condensed matter contexts. It has been particularly
successful in describing few-body scattering in ultracold gases, notably
in the seminal work by Greene and collaborators on ultracold collisions
\citep{Greene1996PRA,Greene2002PRA,Greene2009JPB,Jia2011PRA,Jia2012PRA,Chen2022PRA,Higgins2025PRC}
and Efimov physics \citep{Esry1999RPL,Yujun2011PRL,Jia2012PRL}. While
AHR can also be applied to four-body problems in three dimensions,
doing so typically requires carefully tailored variational basis sets
due to the increased number of degrees of freedom \citep{Mehta2007arXiv,DIncao2009PRA,Rittenhouse2011JPB}.

In contrast, the reduced dimensionality of 1D systems offers computational
advantages, allowing us to apply numerically exact diagonalization
methods using simple B-spline basis functions. Although some adiabatic
hyperspherical potential curves for 1D four-body systems have been
presented in previous studies \citep{Mehta2007arXiv}, scattering
observables such as the dimer-dimer scattering length have not yet
been computed. A key challenge in these calculations is the presence
of sharp avoided crossings in the adiabatic potential curves, which
complicate numerical integration of the coupled channel equations.

To overcome this difficulty, we employ the slow variable discretization
(SVD) method, which enables stable and accurate computation of scattering
observables \citep{Tolstikhin1996JPB,JiaPhDThesis}. In particular,
we focus on a model interaction known as the $\sinh$-$\cosh$ potential,
for which analytical solutions are available for certain parameter
regimes \citep{Sutherland}. We benchmark our numerical results against
these exact solutions, and further extend the analysis to parameter
ranges where no analytical solutions exist.

\section{Theory}

\subsection{Hyperspherical Coordinate}

\begin{figure}

\includegraphics[width=0.98\textwidth]{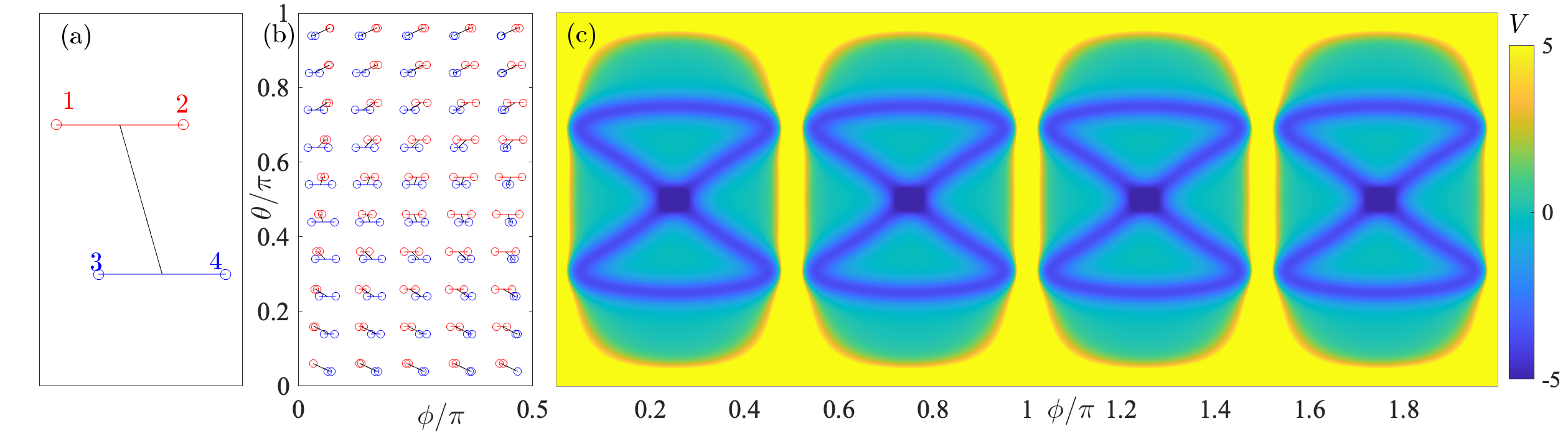}\caption{(a) Schematic illustration of the four-particle configuration. As
the system is one-dimensional, only the horizontal displacement is
physically meaningful; vertical offsets are used solely for visual
clarity. (b) Mapping between the geometric configuration of particles
and the hyperangular coordinates $\theta$ and $\phi$. (c) The four-body
potential $V(R,\Omega)$ plotted as a function of angles at fixed
hyperradius $R=10$. The pairwise two-body interaction is modeled
using the $\sinh$-$\cosh$ potential with parameters $s=s^{\prime}=1.5$.
\protect\label{fig:Sketch}}
\end{figure}

We consider four particles confined to one dimension (1D), with coordinates
$r_{j}$ and masses $m_{j}$ (for $j=1,2,3,4$). It is convenient
to begin by transforming to ``H''-type Jacobi coordinates, where
the configuration is illustrated in Fig. \ref{fig:Sketch} (a). We
define
\begin{equation}
\begin{gathered}y_{1}=\sqrt{\frac{\mu_{1}}{\mu_{{\rm 4b}}}}x_{1}=\sqrt{\frac{\mu_{1}}{\mu_{{\rm 4b}}}}\left(r_{1}-r_{2}\right),\\
y_{2}=\sqrt{\frac{\mu_{2}}{\mu_{{\rm 4b}}}}x_{2}=\sqrt{\frac{\mu_{2}}{\mu_{{\rm 4b}}}}\left(r_{3}-r_{4}\right),\\
y_{3}=\sqrt{\frac{\mu_{3}}{\mu_{{\rm 4b}}}}x_{3}=\sqrt{\frac{\mu_{3}}{\mu_{{\rm 4b}}}}\left(\frac{r_{1}+r_{2}}{2}-\frac{r_{3}+r_{4}}{2}\right),\\
y_{4}=x_{4}=\frac{r_{1}+r_{2}+r_{3}+r_{4}}{4}.
\end{gathered}
\end{equation}
Here, $x_{i}$ are the conventional Jacobi coordinate, while $y_{i}$
are their corresponding mass-scaled forms. The centre-of-mass coordinate
$y_{4}=x_{4}$ can usually be separated from the dynamics. The reduced
masses are defined by $\mu_{1}^{-1}=m_{1}^{-1}+m_{2}^{-1}$, $\mu_{2}^{-1}=m_{3}^{-1}+m_{4}^{-1}$,
and $\mu_{3}^{-1}=\left(m_{1}+m_{2}\right)^{-1}+\left(m_{3}+m_{4}\right)^{-1}$.
The four-body reduced mass $\mu_{{\rm 4b}}$ can essentially be defined
arbitrarily as long as the relative kinetic energy operator are defined
consistently. We choose $\mu_{{\rm 4b}}=\sqrt[3]{\Pi_{j}m_{j}/\left(2\sum_{j}m_{j}\right)}$,
and the relative kinetic energy operator reads as

\begin{equation}
\hat{T}=-\sum_{i=1}^{3}\frac{\hbar^{2}}{2\mu_{i}}\frac{\partial^{2}}{\partial x_{i}^{2}}=-\frac{\hbar^{2}}{2\mu_{{\rm 4b}}}\sum_{i=1}^{3}\frac{\partial^{2}}{\partial y_{i}^{2}}.
\end{equation}
Hereafter, we focus on the equal-mass case: of $m_{j}=m$ for all
$j$. Then $\mu_{{\rm 4b}}=m/2$, $\mu_{1}=m/2$, $\mu_{2}=m/2$,
and $\mu_{3}=m$. We assume particles 1 and 2 are identical fermions,
as are particles 3 and 4, but the two pairs are distinguishable from
each other. The pairwise distances $r_{ij}$ between the particles
$i$ and $j$ become: 

\begin{equation}
\begin{gathered}r_{12}=y_{1},\\
r_{13}=y_{1}/2-y_{2}/2+y_{3}/\sqrt{2},\\
r_{14}=y_{1}/2+y_{2}/2+y_{3}\sqrt{2},\\
r_{23}=-y_{1}/2-y_{2}/2+y_{3}/\sqrt{2},\\
r_{24}=-y_{1}/2+y_{2}/2+y_{3}/\sqrt{2},\\
r_{34}=y_{2}.
\end{gathered}
\end{equation}

We introduce hyperspherical coordinates in analogy with three-dimensional
(3D) spherical coordinate:
\begin{equation}
\begin{gathered}y_{1}=R\sin\phi\sin\theta\\
y_{2}=R\cos\phi\sin\theta\\
y_{3}=R\cos\theta
\end{gathered}
\end{equation}
with the hyperradius $R\in[0,\infty)$, and the hyperangles $\Omega\equiv\{\theta,\phi\}$
where $\phi\in[0,2\pi)$ and $\theta\in[0,\pi]$. Figure \ref{fig:Sketch}
(b) illustrates the mapping between the geometric configuration of
the four particles and the hyperangles. For clarity, we only show
the region $\phi\in[0,\pi/2]$ and $\theta\in[0,\pi]$, as the full
configuration space can be reconstructed using symmetry operations
(see discussion below). 

The relative kinetic energy operator becomes:

\begin{equation}
\hat{T}=-\frac{\hbar^{2}}{2\mu_{{\rm 4b}}}\frac{1}{R}\frac{\partial^{2}}{\partial R^{2}}R+\frac{\hat{L}\left(\Omega\right)^{2}}{2\mu_{{\rm 4b}} R^{2}},
\end{equation}
with the hyperangular momentum operator defined as:

\begin{equation}
\hat{L}\left(\Omega\right)^{2}=-\hbar^{2}\left[\frac{1}{\sin\theta}\frac{\partial}{\partial\theta}\sin\theta\frac{\partial}{\partial\theta}+\frac{1}{\sin^{2}\theta}\frac{\partial^{2}}{\partial\phi^{2}}\right].
\end{equation}
This operator has the same form as the angular momentum operator in
three-dimensional space, which is expected since the number of relative
degrees of freedom in our four-body problem is also three.

The four-body Schr{\"o}dinger equation (after separating the center-of-mass
degree of freedom) is given by

\begin{equation}
\left[-\frac{\hbar^{2}}{2\mu_{{\rm 4b}}}\frac{1}{R}\frac{\partial^{2}}{\partial R^{2}}R+\frac{\hat{L}\left(\Omega\right)^{2}}{2\mu_{{\rm 4b}} R^{2}}+V\left(R,\Omega\right)\right]\Psi\left(R,\Omega\right)=E\Psi\left(R,\Omega\right).
\end{equation}
We define a rescaled wave function $\psi=R\Psi$, which satisfies
\begin{equation}
\left[-\frac{\hbar^{2}}{2\mu_{{\rm 4b}}}\frac{\partial^{2}}{\partial R^{2}}+\frac{\hat{L}\left(\Omega\right)^{2}}{2\mu_{{\rm 4b}} R^{2}}+V\left(R,\Omega\right)\right]\psi\left(R,\Omega\right)=E\psi\left(R,\Omega\right),\label{eq:Sch4b}
\end{equation}
where $V\left(R,\Omega\right)$ denotes the interaction potential
among particles.

\subsection{The $\sinh$-$\cosh$ potential}

For ultracold atomic systems, the total interaction potential can
usually be modeled as a sum of pairwise interactions:
\begin{equation}
V\left(R,\Omega\right)=\sum_{i<j}v_{ij}\left(r_{ij}\right),
\end{equation}
where the interparticle distances $r_{ij}$ can then be explicitly
written in terms of hyperspherical coordinates:
\begin{equation}
\begin{gathered}r_{12}=R\sin\phi\sin\theta,\\
r_{13}=R\left[\sin\theta\sin\left(\phi-\pi/4\right)+\cos\theta\right]/\sqrt{2},\\
r_{14}=R\left[\sin\theta\sin\left(\phi+\pi/4\right)+\cos\theta\right]/\sqrt{2},\\
r_{23}=-R\left[\sin\theta\sin\left(\phi+\pi/4\right)-\cos\theta\right]/\sqrt{2},\\
r_{24}=-R\left[\sin\theta\sin\left(\phi-\pi/4\right)-\cos\theta\right]/\sqrt{2},\\
r_{34}=R\cos\phi\sin\theta.
\end{gathered}
\label{eq:rij_hyper}
\end{equation}

The system we now wish to focus consists of two kinds of fermions
with the same mass $m$, distinguished only by a label $\sigma=\pm1$.
The pair-wise potential considered here is called the $\sinh$-$\cosh$
potential \cite{Sutherland}, which follows the form of
\begin{equation}
v_{\sigma_{i}\sigma_{j}}(r_{ij})=\frac{s^{\prime}\left(s^{\prime}+1\right)}{\sinh^{2}\left(r_{ij}/r_{c}\right)}\left(\frac{1+\sigma_{i}\sigma_{j}}{2}\right)-\frac{s\left(s+1\right)}{\cosh^{2}\left(r_{ij}/r_{c}\right)}\left(\frac{1-\sigma_{i}\sigma_{j}}{2}\right),
\end{equation}
where $s^{\prime},s>0$. This form ensures that like particles (same
$\sigma$) interact repulsively, while unlike particles experience
an attractive interaction. In this sense, the system is analogous
to an electron-hole model, where $\sigma$ can be interpreted as an
effective charge. At large distances, both types of interactions decay
exponentially with a characteristic length scale $r_{c}$, which we
hereafter set to unity. The corresponding natural energy scale is
then $E_c = \hbar^{2}/mr_{c}^{2}$. Throughout this work, we adopt
$r_c$ and $E_c$ as natural units to interpret the length and energy scales
in the figures.

Figure \ref{fig:Sketch} (c) shows the four-body potential at hyperradius
$R=10$ as a function of the hyperangles. The plot reveals a clear
fourfold symmetry, with identical potential landscapes across four
regions. Notably, a deep minimum appears near $\left(\theta,\phi\right)=\left(\pi/2,\pi/4\right)$,
indicated by the dark blue region, while a lighter blue figure-eight-shaped
stripe and a broader light green plateau surround it. As we will see,
these regions correspond to different physical configurations: the
deep minimum supports dimer-dimer scattering states, the lighter stripe
indicates dimer--atom--atom configurations, and the outer plateau
corresponds to four free atoms.

One interesting property of the $\sinh$-$\cosh$ potential is that
the two-body scattering properties are known analytically \cite{Sutherland}.
For the repulsive $\sinh$-potential between like particles, with
the relative momentum defined as $k=k_{1}-k_{2}$, the scattering
amplitude is given as

\begin{equation}
S\left(k\right)=-\frac{\Gamma\left(1+ik/2\right)\Gamma\left(s+1-ik/2\right)}{\Gamma\left(1-ik/2\right)\Gamma\left(s+1+ik/2\right)},
\end{equation}
where $\Gamma\left(\cdot\right)$ is the gamma function. For unlike
particles, the $\cosh$-potential are attractive, whose reflection
and transmission amplitudes are given as
\begin{equation}
R\left(k\right)=S\left(k\right)r\left(k\right),\ T\left(k\right)=S\left(k\right)t\left(k\right),
\end{equation}
where
\begin{equation}
\begin{gathered}r\left(k\right)=\frac{\sin\pi\left(s+1\right)}{\sin\pi\left(s+1+ik/2\right)},\\
t\left(k\right)=\frac{\sin\pi ik/2}{\sin\pi\left(s+1+ik/2\right)}.
\end{gathered}
\end{equation}
Dimer bound states appear as poles of the reflection and transmission
amplitudes, with binding energy $E_{\tau}=-\left(s+1-\tau\right)^{2}$,
$\tau=1,2,\cdots,\left\lfloor s+1\right\rfloor $, where $\left\lfloor x\right\rfloor $
denote the largest integer less than $x$. The dimer can thus be described
by two particles with momentums $k\pm i\kappa_{\tau}$, where $\kappa_{\tau}=s+1-\tau$. 

For the case of $s=s^{\prime}$, the system become integrable for arbitrary
number of particles. No other bound state exists other than the dimer
bound states mentioned above. The atom-dimer scattering amplitude
between a particle with momentum $k_{1}$ and a pair of bounded particles
with momenta $k_{2}\pm i\kappa_{\tau}$ is given by
\begin{equation}
\exp\left[-i\theta_{0\tau}(k)\right]=S\left(k+i\kappa_{\tau}\right)S\left(k-i\kappa_{\tau}\right)t\left(k+i\kappa_{\tau}\right)
\end{equation}
with $k=k_{2}-k_{1}$. The dimer-dimer scattering amplitude $S_{\tau\tau^{\prime}}(k)=\exp\left[-i\theta_{\tau\tau^{\prime}}\left(k\right)\right]$
between bound atom pairs with $k_{1}\pm i\kappa_{\tau}$ and $k_{2}\pm i\kappa_{\tau^{\prime}}$
are determined by the phase shift 
\begin{equation}
\theta_{\tau\tau^{\prime}}(k)=\theta_{0\tau^{\prime}}(k-i\kappa_{\tau})+\theta_{0\tau^{\prime}}(k+i\kappa_{\tau}),
\end{equation}
with $k=k_{2}-k_{1}$. For scattering between the ground state dimers
$\tau=\tau^{\prime}=1$, we have $\kappa_{1}=s$, and $E_{1}=-s^{2}$.
For $s\ge1$, the ground state dimer-dimer scattering amplitude has
a simplified analytical form
\begin{equation}
\mathcal{S}(k)=-\frac{\left(s-ik/2\right)}{\left(s+ik/2\right)}\frac{\Gamma\left(1+ik/2\right)}{\Gamma\left(1-ik/2\right)}\frac{\Gamma\left(2s+1-ik/2\right)}{\Gamma\left(2s+1+ik/2\right)}.
\end{equation}
Since this is the lowest scattering channel, this scattering amplitude
is equivalent to the $S$-matrix. The closely related $K$-matrix
is given by 
\begin{equation}
\mathcal{K}(k)=i\frac{1-\mathcal{S}\left(k\right)}{1+\mathcal{S}\left(k\right)},
\end{equation}
and the energy-dependent dimer-dimer scattering length is given by
\begin{equation}
a_{D}(k)=\frac{1}{k\mathcal{K}(k)}.\label{eq:ak}
\end{equation}
The dimer-dimer scattering length at zero scattering energy is thus
given by
\begin{equation}
a_{D}\equiv\lim_{k\rightarrow0}a_{D}\left(k\right)=\frac{\gamma+s^{-1}+\psi\left(2s+1\right)}{2},
\end{equation}
where $\psi\left(\cdot\right)$ is the digamma function and $\gamma=0.577216$
is the Euler-Mascheroni constant. We will use these analytical results
to benchmark our numerical calculations.

\section{Results}

\subsection{Hyperspherical potential}

To efficiently solve Eq. (\ref{eq:Sch4b}), we adopt the AHR. This involves
solving the hyperangular eigenvalue problem:
\begin{equation}
\left[\frac{\hat{L}\left(\Omega\right)^{2}}{2\mu_{{\rm 4b}} R^{2}}+V\left(R,\Omega\right)\right]\Phi_{\nu}(R;\Omega)=U_{\nu}\left(R\right)\Phi_{\nu}\left(R;\Omega\right),
\end{equation}
where $U_{\nu}\left(R\right)$ are the hyperspherical potentials,
and $\Phi_{\nu}(\Omega;R)$ are the corresponding channel functions,
which depend parametrically on hyperradius $R$. These functions form
a truncated basis for expanding the total wave function:
\begin{equation}
\psi_{\nu^{\prime}}^{E}(R,\Omega)=\sum_{\nu=1}^{N_{c}}F_{\nu\nu^{\prime}}(R)\Phi_{\nu}(\Omega;R),
\end{equation}
where $\nu^{\prime}$ indexes the independent solutions and $N_{c}$
is the number of channels retained in the expansion.

Before proceeding further, we note that the calculation can often
be simplified by imposing permutation symmetry for identical particles.
In what follows, we focus on the case of four fermions, where particles
1 and 2 are assigned $\sigma=1$, and particles 3 and 4 are assigned
$\sigma=-1$. The total wave function must be antisymmetrized using
the operator
\begin{equation}
\mathcal{A}=1-\mathcal{P}_{12}-\mathcal{P}_{34}+\mathcal{P}_{12}\mathcal{P}_{34},
\end{equation}
where $P_{ij}$ denotes the exchange operator for particles $i$ and
$j$. From the expressions of the interparticle distances $r_{ij}$
in Eq. (\ref{eq:rij_hyper}), we find that $P_{12}$ pas $\phi\rightarrow2\pi-\phi$,
while $P_{34}$ maps $\phi\rightarrow\pi-\phi$, revealing the fourfold
symmetry observed in the potential landscape shown in Fig. \ref{fig:Sketch}
(c). Since the hyperangular momentum operator takes the same form
as the angular momentum operator for a single particle in three-dimensional
space, the wave function can be formally expanded using spherical
harmonics: $u_{\ell m_{\ell}}\left(\theta,\phi\right)\sim P_{\ell}^{m_{\ell}}\left(\cos\theta\right)e^{im_{\ell}\phi}$.
Applying the antisymmetrization operator $\mathcal{A}$ to these functions
yields antisymmetrized basis functions of the form $u_{\ell k_{\ell}}^{\mathcal{A}}\left(\theta,\phi\right)\sim P_{\ell}^{2k_{\ell}}\left(\cos\theta\right)\sin\left(2k_{\ell}\phi\right)$,
where $k_{\ell}$ is a positive integers. 

The pair-wise potential $v_{\sigma_{i}\sigma_{j}}(r)$ also exhibits
certain symmetries: it is even in $r$, i.e., $v_{\sigma_{i}\sigma_{j}}(r)=v_{\sigma_{i}\sigma_{j}}(-r)$,
and identical for like-particle pairs, i.e., $v_{11}\left(r\right)=v_{-1-1}\left(r\right)$.
As a result, the full Hamiltonian is invariant under two commuting
operations $\mathcal{P}_{\theta}:\theta\rightarrow\pi-\theta$ and
$\mathcal{P}_{\phi}:\phi\rightarrow\pi/2-\phi$. These symmetries
allow us to define two good quantum numbers: $p_{\theta}=\left(-\right)^{\ell}$,
$p_{\phi}=\left(-\right)^{k_{\ell}+1}$. Consequently, the calculation
can be restricted to the reduced domain $\theta\in[0,\pi/2]$ and
$\phi\in[0,\pi/4]$, with numerical basis $u_{i}\left(\theta,\phi\right)$
subject to the appropreate boundary conditions. We find that the ground-state
dimer-dimer channel lies in the symmetry sector with $p_{\theta}=+1$
and $p_{\phi}=+1$, corresponding to even $\ell$ and odd $k_{\ell}$,
which implies $u_{i}\left|_{\phi=0}\right.=0$, $\partial_{\phi}u_{i}\left|_{\phi=\pi/4}\right.=0$,
$\partial_{\theta}u_{i}\left|_{\theta=\pi/2}\right.=0$ and $u_{i}\left|_{\theta=0}\right.$
remains finite. (Boundary conditions for other symmetry sector are
summarized in Table \ref{tab:BC}.) We employ B-spline basis functions, which are piecewise polynomials with local support. This choice facilitates the implementation of boundary conditions and leads to a sparse Hamiltonian matrix that can be efficiently diagonalized using eigensolvers such as ARPACK. In practice, we typically use approximately $100$ B-splines for the $\theta$ coordinate and 50 for $\phi$, resulting in a sparse matrix of dimension around $5,000$.

\begin{table}[h]
\caption{Boundary conditions for the numerical basis functions in different
symmetry sectors, characterized by $p_{\theta}$ and $p_{\phi}$.}\label{tab:BC}

\begin{tabular}{@{}lllllll@{}}
\toprule
$p_{\theta}$ & $p_{\phi}$ & $\theta=0$ & $\theta=\pi/2$ & $\phi=\pi/4$ & $\phi=\pi/4$ \\
\midrule
$+1$ & $+1$ & Finite & $\partial_{\theta}u_{i}=0$ & $u_{i}=0$ & $\partial_{\phi}u_{i}=0$ \\ 
$+1$ & $-1$ & Finite & $\partial_{\theta}u_{i}=0$ & $u_{i}=0$ & $u_{i}=0$ \\
$-1$ & $+1$ & Finite & $u_{i}=0$ & $u_{i}=0$ & $\partial_{\phi}u_{i}=0$ \\
$-1$ & $-1$ & Finite & $u_{i}=0$ & $u_{i}=0$ & $u_{i}=0$ \\
\botrule 
\end{tabular}
\end{table}

\begin{figure}
\includegraphics[width=0.98\textwidth]{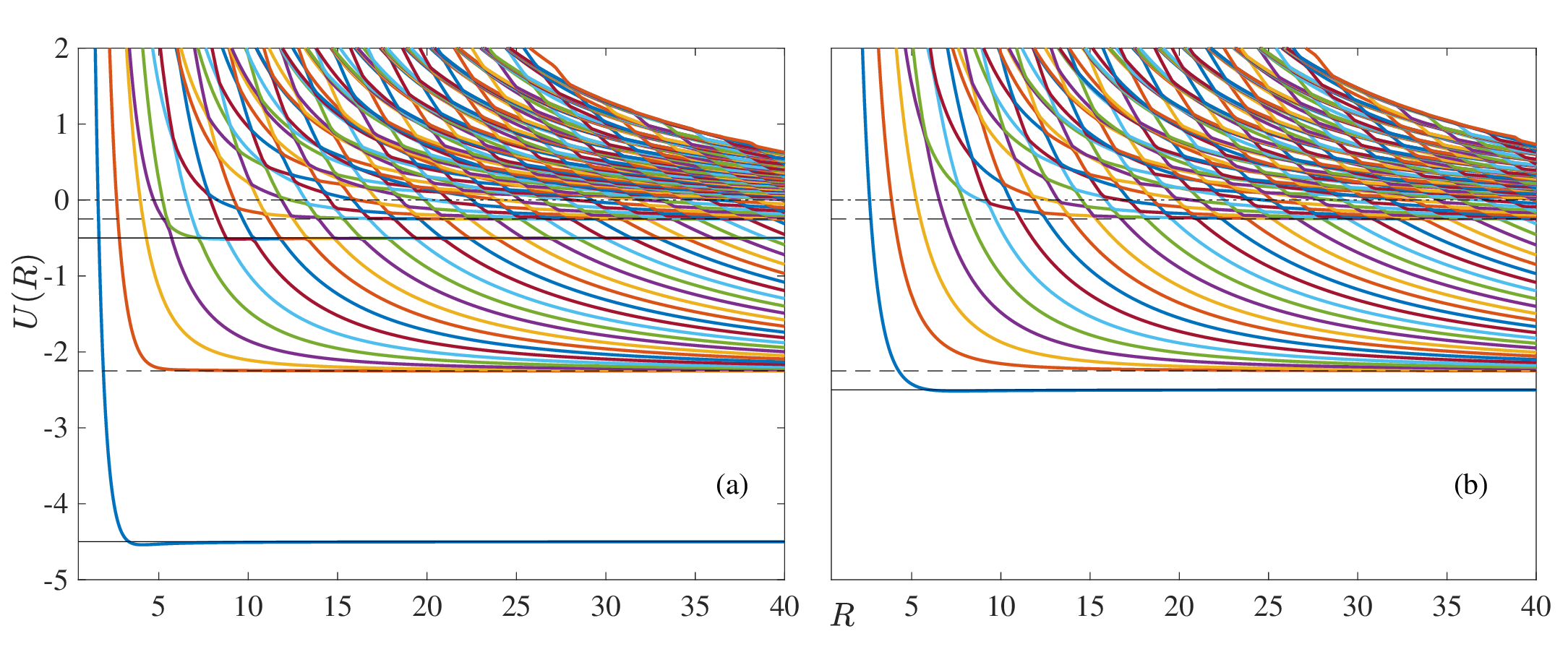}\caption{Hyperspherical potential curves for $s=s^{\prime}=1.5$ in symmetry
sectors: (a) $\left(p_{\theta},p_{\phi}\right)=\left(+1,+1\right)$
and (b) $\left(p_{\theta},p_{\phi}\right)=\left(+1,-1\right)$. Solid
horizontal lines indicate dimer-dimer thresholds; dashed lines mark
dimer--atom--atom thresholds; dash-dotted lines represent the atom-atom-atom-atom
threshold at zero energy.\protect\label{fig:UR1}}
\end{figure}

\begin{figure}
\includegraphics[width=0.98\textwidth]{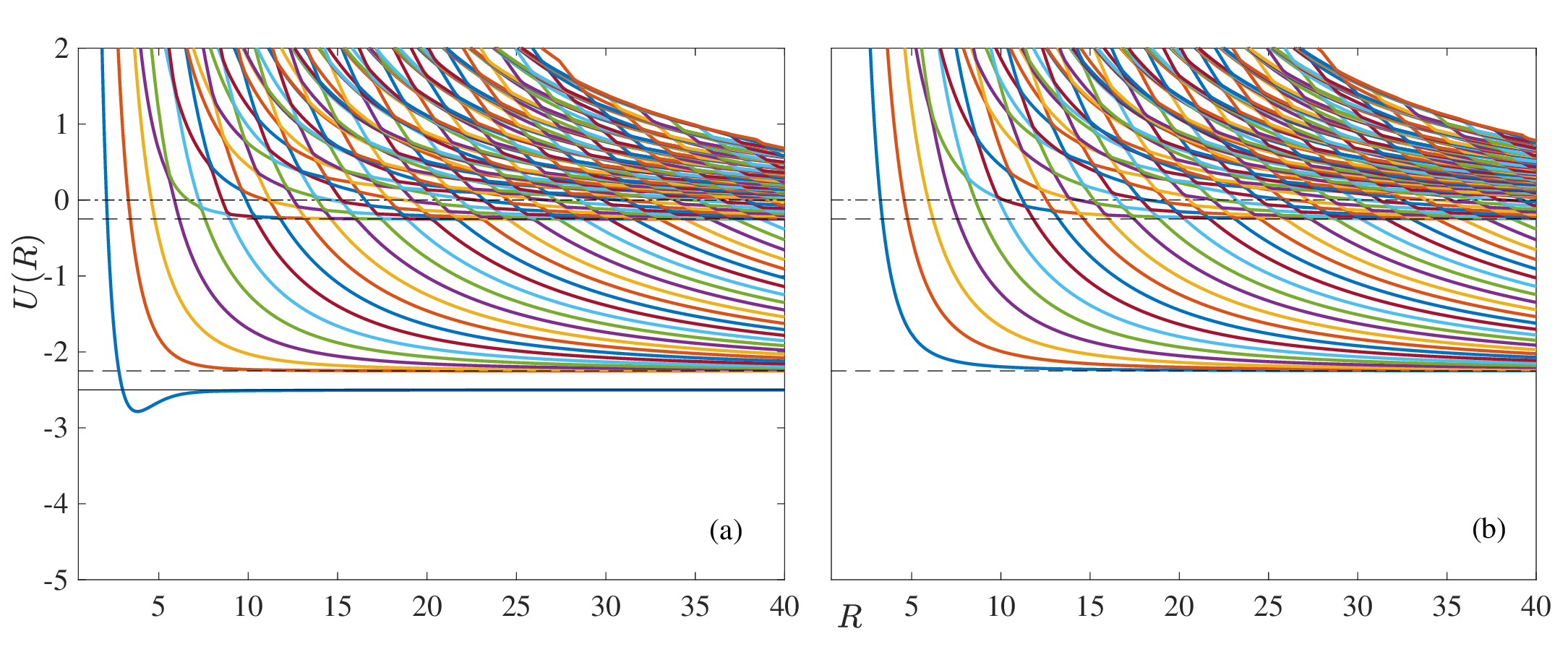}\caption{Same as Fig. \ref{fig:UR1}, but for symmetry sectors: (a) $\left(p_{\theta},p_{\phi}\right)=\left(-1,+1\right)$
and (b) $\left(p_{\theta},p_{\phi}\right)=\left(-1,-1\right)$. \protect\label{fig:UR2}}
\end{figure}

\begin{figure}
\includegraphics[width=0.98\textwidth]{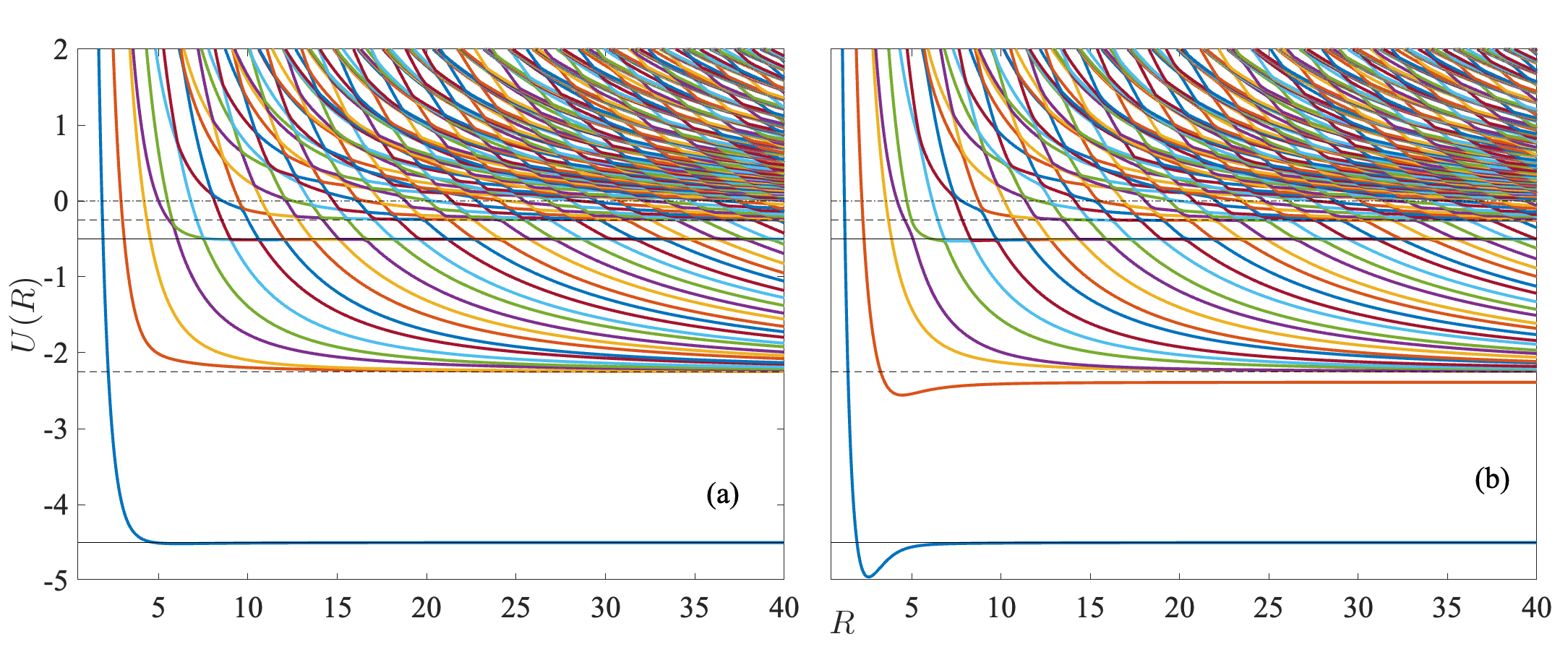}\caption{Hyperspherical potential curves for $\left(p_{\theta},p_{\phi}\right)=\left(+1,+1\right)$ and $s=1.5$ with different $s^{\prime}$:
(a) $s^{\prime}=1.9$ and (b) $s^{\prime}=0.9$. In panel (b), the
second-lowest channel approaches a trimer--atom threshold due to
weakened repulsion between like particles, indicating trimer formation.
\protect\label{fig:URp}}
\end{figure}

Figure \ref{fig:UR1} and \ref{fig:UR2} present the hyperspherical
potentials for $s=s^{\prime}=1.5$ across different symmetry sectors,
while Figure \ref{fig:URp} shows the hyperspherical potentials for
fixed $s=1.5$ and varying $s^{\prime}$. A common feature among these
curves is their convergence toward specific asymptotic thresholds
at large hyperradii. For $s=1.5$, the attractive interaction between
unlike particles supports two dimer bound states with binding energies
$E_{1}=-2.25$ and $E_{2}=-0.25$. As a result, the dimer--dimer
thresholds appear at $E_{1}+E_{1}=-4.5$, $E_{1}+E_{2}=-2.5$, and
$E_{2}+E_{2}=-0.5$, indicated by solid horizontal lines in the figures.
The dimer--atom--atom thresholds, corresponding to single dimer
binding energies $E_{1}$ and $E_{2}$, are shown as dashed horizontal
lines. The atom--atom--atom--atom threshold lies at zero energy
and is indicated by dash-dotted lines. An exception occurs in the
second-lowest channel of Fig. \ref{fig:URp} (b), which does not approach
any of the aforementioned thresholds. This behavior arises because
the reduced repulsion between like particles at $s^{\prime}=0.9$
permits the formation of trimer bound states, causing this channel
to asymptotically approach a trimer--atom threshold. Additionally,
many sharp avoided crossings are observed between potential curves
that approach different thresholds, which may pose numerical challenges
for accurate scattering calculations.

\begin{figure}
\includegraphics[width=0.98\textwidth]{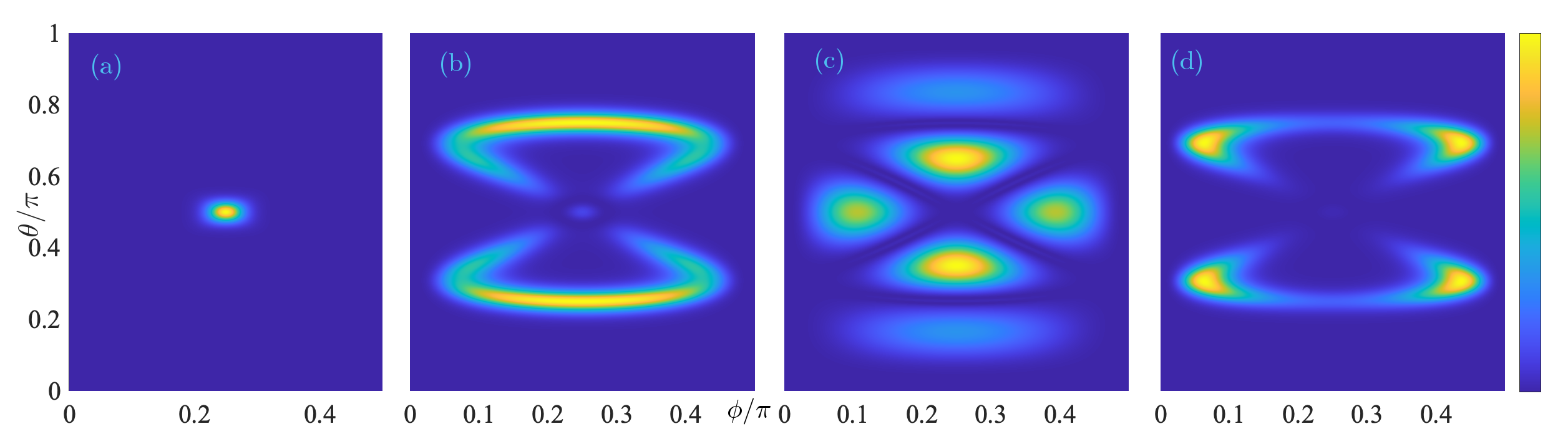}

\caption{Channel functions for (a) dimer-dimer channel (b) dimer-atom-atom
channel (c) Four-atom channel and (d) trimer-atom channel. The color
scale indicates the probability density of the wave function, shown
in arbitrary units. \protect\label{fig:CFun}}
\end{figure}

Figure \ref{fig:CFun} displays the channel functions $|\Phi_{\nu}(\Omega;R)|^{2}$
for several representative configurations. In panel (a), we show the
dimer--dimer channel function corresponding to the lowest hyperspherical
potential at $R=10$ in Fig. \ref{fig:UR1} (a). Panel (b) shows the
dimer--atom--atom channel function from the second-lowest channel
at the same $R=10$ in Fig. \ref{fig:UR1} (a). Panel (c) presents
the atom--atom--atom--atom channel function from the 41st channel
at $R=40$ in Fig. \ref{fig:UR1} (a). Finally, panel (d) shows the
trimer--atom channel function corresponding to the lowest potential
at $R=10$ in Fig. \ref{fig:URp} (b). These channel functions reveal
the spatial structure of the associated scattering configurations.
The dimer--dimer channel is primarily localized in the deep blue
minimum of the four-body potential landscape {[}see Fig. \ref{fig:Sketch}
(c){]}, while the atom--atom--atom--atom configuration occupies
the broader light green plateau. The dimer--atom--atom and trimer--atom
configurations are mostly concentrated along the figure-eight-shaped
light blue region, highlighting the distinct geometrical signatures
of different few-body scattering processes.


\subsection{Scattering}

With the adiabatic potential curves and diabatic corrections obtained
from the channel functions, we solve the radial Schr{\"o}dinger equation
as a set of coupled differential equations. However, near avoided
crossings, the diabatic corrections can exhibit rapid variations that
lead to numerical instability. To address this challenge, we use the
slow variable discretization (SVD) method with a radial discrete variable
representation (DVR) basis $\pi_{n}(R)$ to expand $F_{\nu\nu^{\prime}}^ {}(R)$,
which gives 
\begin{equation}
\psi_{\nu^{\prime}}^ {}(R,\Omega)=\sum_{\nu,n}c_{n\nu,\nu^{\prime}}\pi_{n}(R)\Phi_{\nu}\left(\Omega;R_{n}\right).
\end{equation}
The DVR approximation allows integrals of the form:

\begin{equation}
\int_{a_{1}}^{a_{2}}\pi_{i}(R)H(R)\pi_{j}(R)dR\cong H\left(R_{i}\right)\delta_{ij}
\end{equation}
for integrating smooth function $H(R)$ with quadrature method. Under
this approximation, the expansion coefficients $c_{n\nu,\nu^{\prime}}$
satisfy the following equation:
\begin{equation}
\sum_{n,\mu}T_{nn^{\prime}}O_{n\nu,n^{\prime}\mu}c_{n^{\prime}\mu,\nu^{\prime}}+\left[U_{\nu}\left(R_{n}\right)-E\right]c_{n\nu,\nu^{\prime}}=0,
\end{equation}
with kinetic energy matrix elements: 
\begin{equation}
T_{nn^{\prime}}=\int dR\pi_{n}(R)\left[-\frac{1}{2\mu_{{\rm 4b}}}\frac{\partial^{2}}{\partial R^{2}}\right]\pi_{n^{\prime}}(R)dR
\end{equation}
and overlap matrix:
\begin{equation}
O_{n\nu,n^{\prime}\mu}=\int d\Omega\Phi_{\nu}\left(\Omega;R_{n}\right)^{*}\Phi_{\mu}\left(\Omega;R_{n^{\prime}}\right).
\end{equation}
Once the expansion coefficients $c_{n\nu,\nu^{\prime}}$ are determined,
the total wave function is fully specified, allowing access to all
relevant information for quantum scattering analysis.

In practice, we compute the $R$-matrix, which serves a similar role
to the inverse logarithmic derivative of the wave function in single-channel
scattering problems. It is defined as
\begin{equation}
\mathcal R(R)=F(R)[\tilde{F}(R)]^{-1},
\end{equation}
where $F(R)$ and $\tilde{F}(R)$ denote the wave function and its
derivative, respectively, evaluated at a given hyperradius $R$. At
large distances, the physical scattering matrix $\mathcal{S}$ (and
the related reaction matrix $\mathcal{K}$) can be obtained by applying
asymptotic boundary conditions. For dimer-dimer scattering in the
lowest channel, we use:

\begin{equation}
\ensuremath{\begin{aligned}\mathcal{K} & =\left(f-f^{\prime}\mathcal{R}\right)\left(g-g^{\prime}\mathcal{R}\right)^{-1},\end{aligned}
}
\end{equation}
\begin{equation}
\mathcal{S}=(1+i\mathcal{K})(1-i\mathcal{K})^{-1},
\end{equation}
where $f$ and $g$are the regular and irregular solutions in the
lowest symmetric channel:
\begin{equation}
f=\sqrt{\frac{2\mu_{4b}}{\pi k_{H}}}\cos\left(k_{H}R\right)
\end{equation}
\begin{equation}
g=\sqrt{\frac{2\mu_{4b}}{\pi k_{H}}}\sin\left(k_{H}R\right)
\end{equation}
and $k_{H}=\sqrt{2\mu_{4b}\left(E-E_{11}\right)}$, which is the hyperradial
momentum with respect to the dimer--dimer threshold energy $E_{11}=E_{1}+E_{1}$.
The scattering energy between two dimers is then $E_{s}=E-E_{11}$, 
and the dimer-dimer relative momentum is
$k=\sqrt{2}k_{H}$. The energy-dependent scattering length is computed
using Eq. (\ref{eq:ak}). 

\begin{figure}
\includegraphics[width=0.98\textwidth]{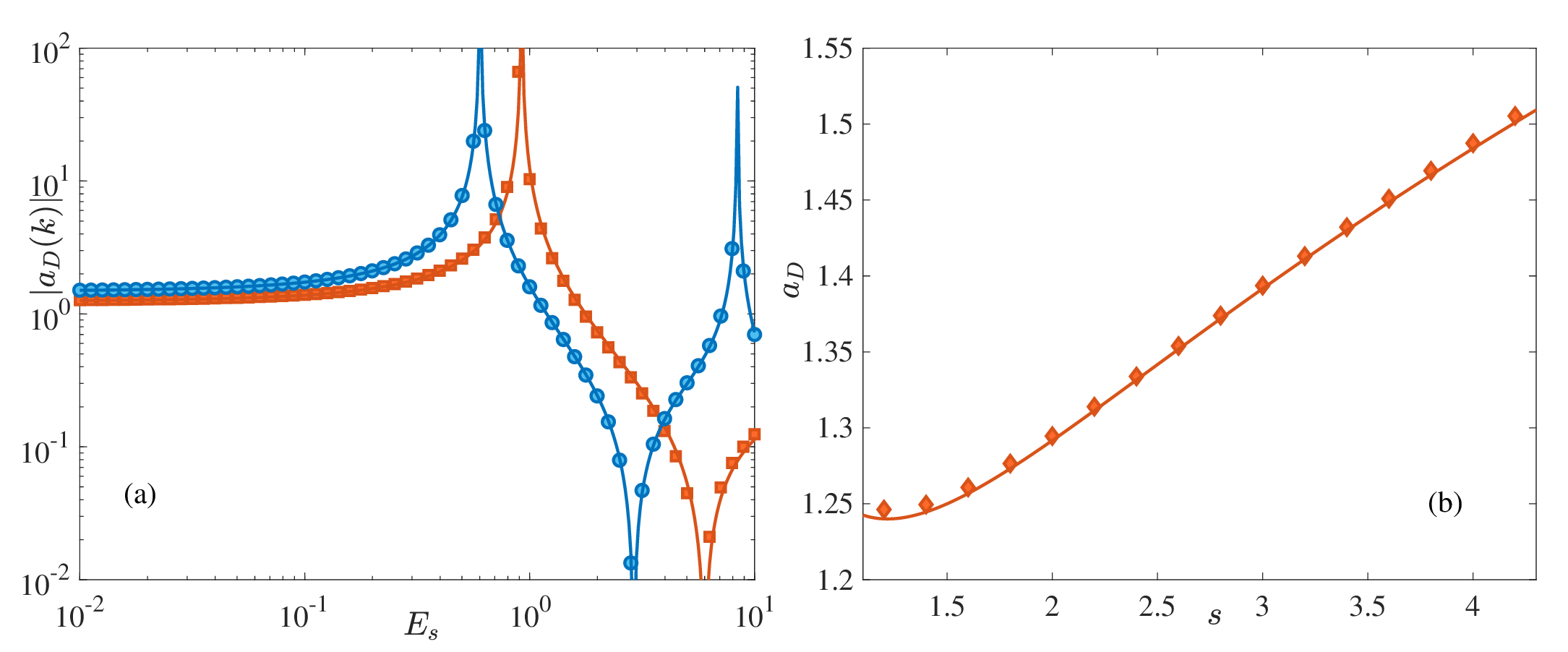}

\caption{Dimer--dimer scattering length between ground-state dimers for the
case $s=s^{\prime}$. (a) Numerical results for the energy-dependent
scattering length $a_{D}(k)$ as a function of scattering energy $E_{s}$,
shown for $s=1.5$ (red squares) and $s=4$ (blue circles). Solid
curves represent the corresponding analytical solutions. (b) Zero-energy
scattering length $a_{D}$ as a function of $s$. Red diamonds indicate
numerical results; the solid curve shows the analytical prediction.\protect\label{fig:aD_ss}}
\end{figure}

\begin{figure}
\includegraphics[width=0.98\textwidth]{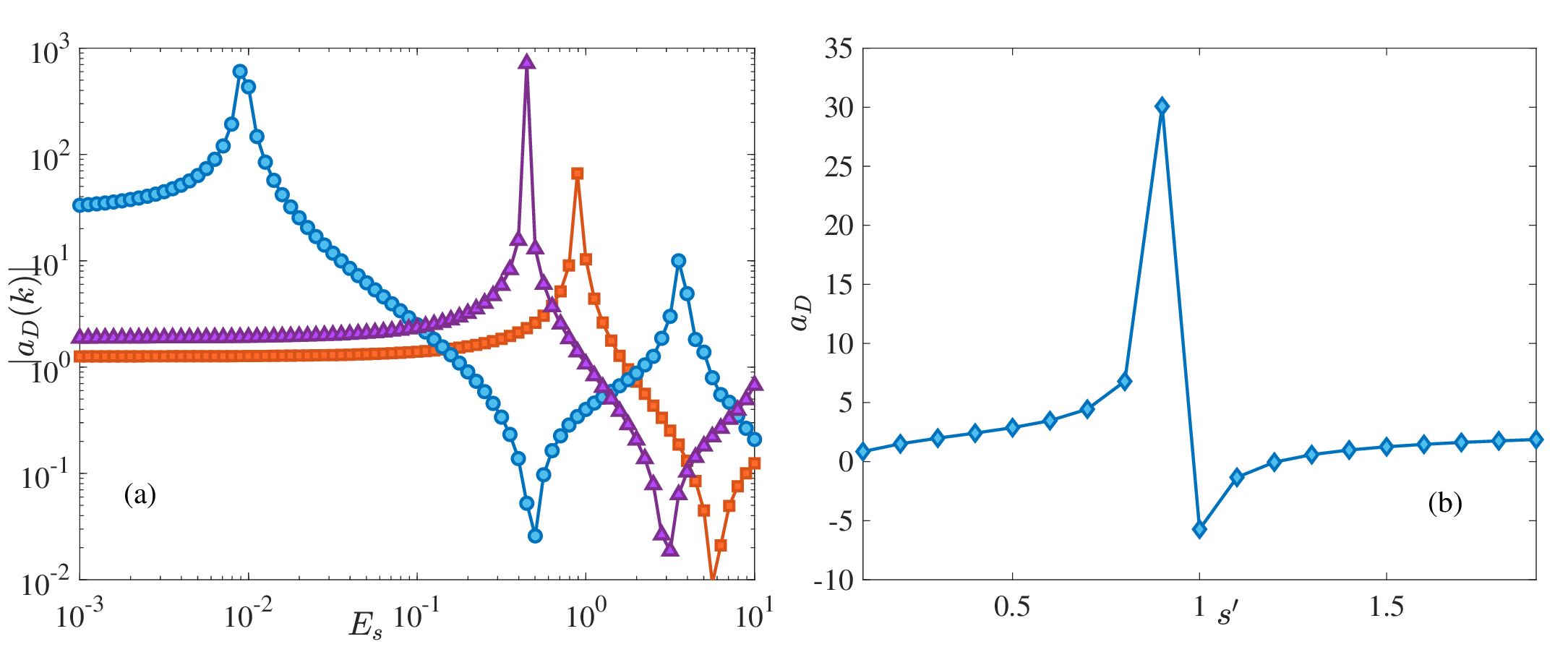}

\caption{Dimer--dimer scattering length between ground-state dimers for $s=1.5$
and varying $s^{\prime}$. Solid curves are provided as visual guides;
analytical results are not available for $s\protect\ne s^{\prime}$.
(a) Energy-dependent scattering length $a_{D}(k)$ as a function of
scattering energy $E_{s}$, for $s^{\prime}=0.9$ (blue circles),
$1.5$ (red squares), and $1.9$ (purple triangles). (b) Zero-energy
scattering length $a_{D}$ as a function of $s^{\prime}$, computed
numerically. \protect\label{fig:aD_sp}}
\end{figure}

Figures \ref{fig:aD_ss} and \ref{fig:aD_sp} present the computed
dimer--dimer scattering lengths between ground-state dimers of the
$\sinh$-potential. In Fig. \ref{fig:aD_ss} (a), the energy-dependent
scattering length $a_{D}(k)$ is shown as a function of scattering
energy $E_{s}$ for $s=s^{\prime}=1.5$ (red squares) and $s=s^{\prime}=4$
(blue circles). The solid curves represent the corresponding analytical
solutions, with excellent agreement across all energies. Panel (b)
shows the zero-energy scattering length as a function of $s$, numerical
results (red diamonds) again match the analytical curve closely. In
contrast, Fig. \ref{fig:aD_sp} explores the more general case where
$s\ne s^{\prime}$, for a fixed $s$ and varying $s^{\prime}$. In
panel (a), the energy-dependent scattering length is plotted $s^{\prime}=0.9$
(blue circles), $1.5$ (red squares), and $1.9$ (purple triangles).
As $s^{\prime}$ decreases, the repulsion between like particles weakens,
allowing the formation of a tetramer. When the tetramer becomes bound,
a resonance appears in the scattering length, which diverges as shown
in panel (b). This resonance occurs near $s^{\prime}\approx0.9$,
signaling the onset of tetramer binding.


\section{Summary and Extension}

In this work, we developed a robust and numerically accurate framework for solving four-body scattering in one-dimensional quantum systems using the adiabatic hyperspherical representation in combination with the slow-variable discretization method. This approach enables precise determination of dimer-dimer scattering amplitudes, even in the presence of sharp avoided crossings in the hyperspherical potentials.

We validated our method against exact analytical results available for the $\sinh$-$\cosh$ interaction model, obtaining excellent agreement across a wide range of energies.  Extending beyond the analytically tractable regime, we explored non-integrable parameter regions and identified a resonance in the dimer-dimer scattering length that signals the emergence of a tetramer bound state. These results highlight both the accuracy of our approach and its capability to reveal emergent few-body phenomena in low-dimensional systems.

Although the $\sinh$-$\cosh$ potential is chosen primarily for analytical convenience, its short-range character enables access to the universal low-energy regime, where scattering properties depend only on the two-body scattering length and not on microscopic interaction details -- just as in higher-dimensional systems \cite{Petrov2004PRL,Leyronas2006PRA}. In the Appendix \ref{secA1}, we show explicitly that the dimer-dimer scattering length in this model obeys the universal relation $a_{\rm dd} = 0.5 a_{\rm aa, S}$, consistent with known results in Ref. \cite{Gogolin2005PRL,Petrov2018PRA}. Crucially, the $\sinh$-$\cosh$ model also allows us to analytically extract the scattering properties of antisymmetric dimers, and we find that they obey the same universal law, $a_{\rm dd} = 0.5 a_{\rm aa,A}$. A full numerical confirmation of universality would require propagation to much larger hyperradii, where the adiabatic potentials and couplings are already smooth. In that asymptotic regime, a sparse-grid representation combined with interpolation or extrapolation could be employed efficiently and matched to the short-range slow-variable discretization region \cite{JiaPhDThesis}. We defer such large-scale numerical developments to future work.

Our findings pave the way for future investigations into more complex configurations, such as mass-imbalanced mixtures, excited-state collisions, and recombination processes. The hyperspherical approach, enhanced by SVD, offers a powerful platform for quantitatively probing universal physics in ultracold atomic gases.

\backmatter

\bmhead{Acknowledgements}
J. W. acknowledges Nirav Mehta for useful discussions. This research was supported by the Australian Research Council's (ARC) Discovery Program, Grants No. FT230100229 (J. W.), No. DP240101590 (H.H.), and No. DP240100248 (X.-J. L.). 

\begin{appendices}

\section{Universality in the dimer-dimer scattering for the $\sinh$-$\cosh$ model}\label{secA1}
The analytical results in the main text concern dimer-dimer scattering between ground-state dimers of the $\sinh$-$\cosh$ model. Here, in this Appendix, we examine in detail the universal limit associated with the shallowest dimers. A key observation is that the two-body potentials in this model are extremely short-ranged,
\begin{equation}
\frac{s(s+1)}{\sinh^{2}(x)}\rightarrow 4s(s+1)e^{-2x} ,\  \frac{s(s+1)}{\cosh^{2}(x)}\rightarrow -4s(s+1)e^{-2x},
\end{equation}
for $x\rightarrow+\infty$. As in three-dimensional systems  \cite{Petrov2004PRL,Leyronas2006PRA}, this implies that at sufficiently low energies, universality emerges and the scattering properties depend only on the scattering length. 

We first examine the shallowest dimer bound state, which appears as the lowest pole in the transmission and reflection amplitudes, with binding energy 
\begin{equation}
E_{\tau}=-(s+1-\left\lfloor s+1\right\rfloor)^{2} =-(s-\left\lfloor s\right\rfloor)^{2} .
\end{equation} 
One can verify that the shallowest dimer is symmetric (antisymmetric) under $x \rightarrow -x$ when $s$ is even (odd), corresponding respectively to the spin-singlet and spin-triplet sectors.

For a symmetric potential, scattering in the even and odd sectors is decoupled. The scattering matrices in the symmetric and antisymmetric channels are related to the transmission ($T$) and reflection ($R$) amplitudes via
\begin{equation}
\mathcal S_{\rm S} = T+R,\ \mathcal S_{\rm A} = T-R,
\end{equation}
for symmetric and antisymmetric state respectively. At the low energy, we have 
\begin{equation}
\mathcal S_{\rm S} \rightarrow -1 + ik a_{\rm aa, S},\ \mathcal S_{\rm A} \rightarrow 1 - ik a_{\rm aa, A},
\end{equation}
with $k = k_1 - k_2 =2 \sqrt E$ in our notation.

The symmetric and antisymmetric scattering lengths in the $\sinh$-$\cosh$ model are then
\begin{equation}
a_{\rm aa,S}=\gamma+\psi^{(0)}(s+1)+\frac{\pi}{2}\cot(\pi s)+\frac{\pi}{2}\csc(\pi s),
\end{equation}
and
\begin{equation}
a_{\rm aa,A}=\gamma+\psi^{(0)}(s+1)+\frac{\pi}{2}\cot(\pi s)-\frac{\pi}{2}\csc(\pi s),
\end{equation}
where $\gamma$ is the Euler constant and $\psi^{(0)}$ is the digamma function. 

As $s$ is approaching to an integer from the above, $s\rightarrow\left\lfloor s\right\rfloor^+$, we have
\begin{equation}
a_{\rm aa,S}\rightarrow\frac{1}{s-\left\lfloor s\right\rfloor }+H[s],\ a_{\rm aa,A}\rightarrow H[s],
\end{equation}
for even $s$, and
\begin{equation}
\ a_{\rm aa,S}\rightarrow H[s],\ a_{\rm aa,A}\rightarrow\frac{1}{s-\left\lfloor s\right\rfloor }+H[s],
\end{equation}
for odd $s$, where $H[s]=\gamma + \psi(s+1)$ is the harmonic number. To the leading order in $s-\left\lfloor s\right\rfloor$,
\begin{equation}
E_{\left\lfloor s+1\right\rfloor }=-\frac{1}{a_{\rm aa}^{2}}
\end{equation}
with $a_{\rm aa}=a_{\rm aa,S}$ or $a_{\rm aa,A}$ for a symmetric or antisymmetric shallowest dimer, respectively -- recovering the standard two-body universal relation.

Applying the same procedure in the main text to dimer-dimer scattering yields
\begin{equation}
a_{\rm dd} = \frac{1}{2}\pi\cot\pi(s+1) + a_{\rm dd,bg},
\end{equation}
where the (non-universal) background term is
\begin{equation}
2a_{{\rm dd,bg}}=2H\left[s\right]-H\left[s-\left\lfloor s\right\rfloor \right]+H[2s-\left\lfloor s\right\rfloor ]+H\left[\left\lfloor s\right\rfloor \right]-H\left[-\left\lfloor s\right\rfloor \right].
\end{equation}
Thus, near $s \to \lfloor s\rfloor^+$,
\begin{equation}
a_{\rm dd}=\frac{a_{\rm aa}}{2}
\end{equation}
for both symmetric and antisymmetric dimers.

\end{appendices}

\bibliography{Ref4body1D}

\end{document}